\documentclass[aps,reprint,amsmath,amssymb,superscriptaddress]{revtex4-2}

\usepackage{graphicx}
\usepackage{physics} 
\usepackage{amssymb}
\usepackage{pgfplots}
\usepackage{tikz}
\usepackage{pgf}
\usepackage[separate-uncertainty = true]{siunitx}
\DeclareSIUnit\torr{Torr}

\newcommand{\Rb}{\ensuremath{{}^{87}\mathrm{Rb}}}

\pgfplotsset{compat=newest} 
\usepgfplotslibrary{units} 

\begin{document}

\title{A modular optically pumped magnetometer system}
\author{T.~Coussens}
\email{t.coussens@sussex.ac.uk}
\affiliation{Department of Physics and Astronomy, University of Sussex, Brighton, BN1 9QH, United Kingdom}
\affiliation{Physikalisch-Technische Bundesanstalt (PTB), Abbestr. 2-12, 10587 Berlin, Germany}
\author{A.~Gialopsou}
\affiliation{Department of Physics and Astronomy, University of Sussex, Brighton, BN1 9QH, United Kingdom}
\affiliation{Clinical Imaging Sciences Centre, Brighton and Sussex Medical School, University of Sussex, Brighton, BN1 9RR, United Kingdom}
\author{C.~Abel}
\affiliation{Department of Physics and Astronomy, University of Sussex, Brighton, BN1 9QH, United Kingdom}
\author{M.~G.~Bason}
\affiliation{Department of Physics and Astronomy, University of Sussex, Brighton, BN1 9QH, United Kingdom}
\affiliation{RAL Space, UKRI-STFC Rutherford Appleton Laboratory, Didcot OX11 0QX, United Kingdom}
\author{T.~M.~James}
\affiliation{Department of Physics and Astronomy, University of Sussex, Brighton, BN1 9QH, United Kingdom}
\author{W.~Evans}
\affiliation{Physikalisch-Technische Bundesanstalt (PTB), Abbestr. 2-12, 10587 Berlin, Germany}
\author{M.~T.~M.~Woodley}
\affiliation{Department of Physics and Astronomy, University of Sussex, Brighton, BN1 9QH, United Kingdom}
\affiliation{Physikalisch-Technische Bundesanstalt (PTB), Abbestr. 2-12, 10587 Berlin, Germany}
\author{D.~Nightingale}
\affiliation{Department of Physics and Astronomy, University of Sussex, Brighton, BN1 9QH, United Kingdom}
\author{D. Nicolau}
\affiliation{Department of Physics and Astronomy, University of Sussex, Brighton, BN1 9QH, United Kingdom}
\author{L. Page}
\affiliation{Department of Physics and Astronomy, University of Sussex, Brighton, BN1 9QH, United Kingdom}
\author{F.~Oru\v{c}evi\'{c}}
\affiliation{Department of Physics and Astronomy, University of Sussex, Brighton, BN1 9QH, United Kingdom}
\author{P.~Kr\"uger}
\affiliation{Department of Physics and Astronomy, University of Sussex, Brighton, BN1 9QH, United Kingdom}
\affiliation{Physikalisch-Technische Bundesanstalt (PTB), Abbestr. 2-12, 10587 Berlin, Germany}


\begin{abstract}
    To address the demands in healthcare and industrial settings for spatially resolved magnetic imaging, we present a modular optically pumped magnetometer (OPM) system comprising a multi-sensor array of highly sensitive quantum magnetometers.
    This system is designed and built to facilitate fast prototyping and testing of new measurement schemes by enabling quick reconfiguration of the self-contained laser and sensor modules as well as allowing for the construction of various array layouts with a shared light source.
    The modularity of this system facilitates the development of methods for managing high-density arrays for magnetic imaging.
    The magnetometer sensitivity and bandwidth are first characterised in both individual channel and differential gradiometer configurations before testing in a real-world magnetoencephalography environment by measuring alpha rhythms from the brain of a human participant.
    We demonstrate the OPM system in a first-order axial gradiometer configuration with a magnetic field gradient sensitivity of \SI{10}{\femto\tesla/\centi\metre/\sqrt{\hertz}}.
    Bandwidths exceeding \SI{200}{\hertz} were achieved for two independent modules.
    The system's increased temporal resolution allows for the measurement of spinal cord signals, which we demonstrate by using phantom signal trials and comparing with an existing commercial sensor.
    
\end{abstract}

\maketitle

\section{Introduction}
    Since the 1960s, superconducting quantum interference devices (SQUIDs) have been the benchmark for the measurement of ultra-low magnetic fields~\cite{Ryhanen1989-ci,Pizzella2001-pl}.
    However, developments in the spin-exchange relaxation-free (SERF) regime over the past two decades have allowed optically pumped magnetometers (OPMs) to achieve similar sensitivities to SQUIDs \citep{kominis_subfemtotesla_2003}, with a number of distinct advantages.
    SQUIDs require cryogenic cooling which increases costs, limits the portability of the device, and increases the separation between the device and the magnetic source of interest.
    OPMs offer reduced separation distances over SQUIDs, which is of particular interest for magnetoencephalography (MEG)~\cite{Sander2012-kv,Colombo2016-hg,Sheng2017-sp,limes_portable_2020}, where OPM-MEG is able to offer an improved spatial~\cite{Tan1990-yx,Boto2016-ya}, and  spatio-temporal~\cite{gialopsou_improved_2021} resolution.
    
    The advantages of OPMs over SQUIDs also benefit many other applications such as the diagnostic measurement of batteries through non-invasive current mapping~\cite{bason_non-invasive_2022,Hu2020-kn,peng_real-time_2023}, magnetic nanoparticle detection~\cite{johnson_magnetic_2012,everaert_monitoring_2023}, defect analysis in metals~\cite{maddox_through-skin_2022} and for fundamental physics research~\cite{Pustelny2013-dq,Abel2019-tm}.
    However, a number of challenges remain, especially around the construction of dense arrays of OPMs.
    Commercially available OPMs currently focus on single-sensor designs; when multiple units are operated in close proximity, issues arise from radio frequency (RF) cross-talk~\cite{nardelli_reducing_2019}, cross-axis projection errors (CAPE)~\cite{borna_cross-axis_2022}, as well as from high surface temperatures of the sensors.
    To achieve dense OPM array systems, these problems must be addressed.

    In addition, in medical applications there is increasing interest in understanding high-frequency biomagnetic phenomena in MEG~\cite{iivanainen_potential_2020}, magnetospinography (MSG)~\cite{chander_non-invasive_2022} and magnetomyography (MMG)~\cite{marquetand_optically_2021}. 
    Existing OPM-MEG is limited to frequencies up to approximately 100 Hz whereas SQUIDs are able to achieve far larger bandwidths~\cite{drung_low-noise_2006}. 
    Further developments in OPMs are therefore needed in order to apply their improved spatial resolution to the higher temporal frequency domain.

    Here, we present a modular magnetometer system designed from the ground up for inherent scalability while maintaining high performance, allowing for easy adaptability of sensor locations and of sensing regimes. The layout of the magnetometer system is determined by assembling individual modules each with specific functionality. 
    Additionally, optical elements inside these modules may be replaced or reconfigured to allow implementation of different sensing schemes.
    
    \begin{figure}
        \centering
        \includegraphics[width=\columnwidth]{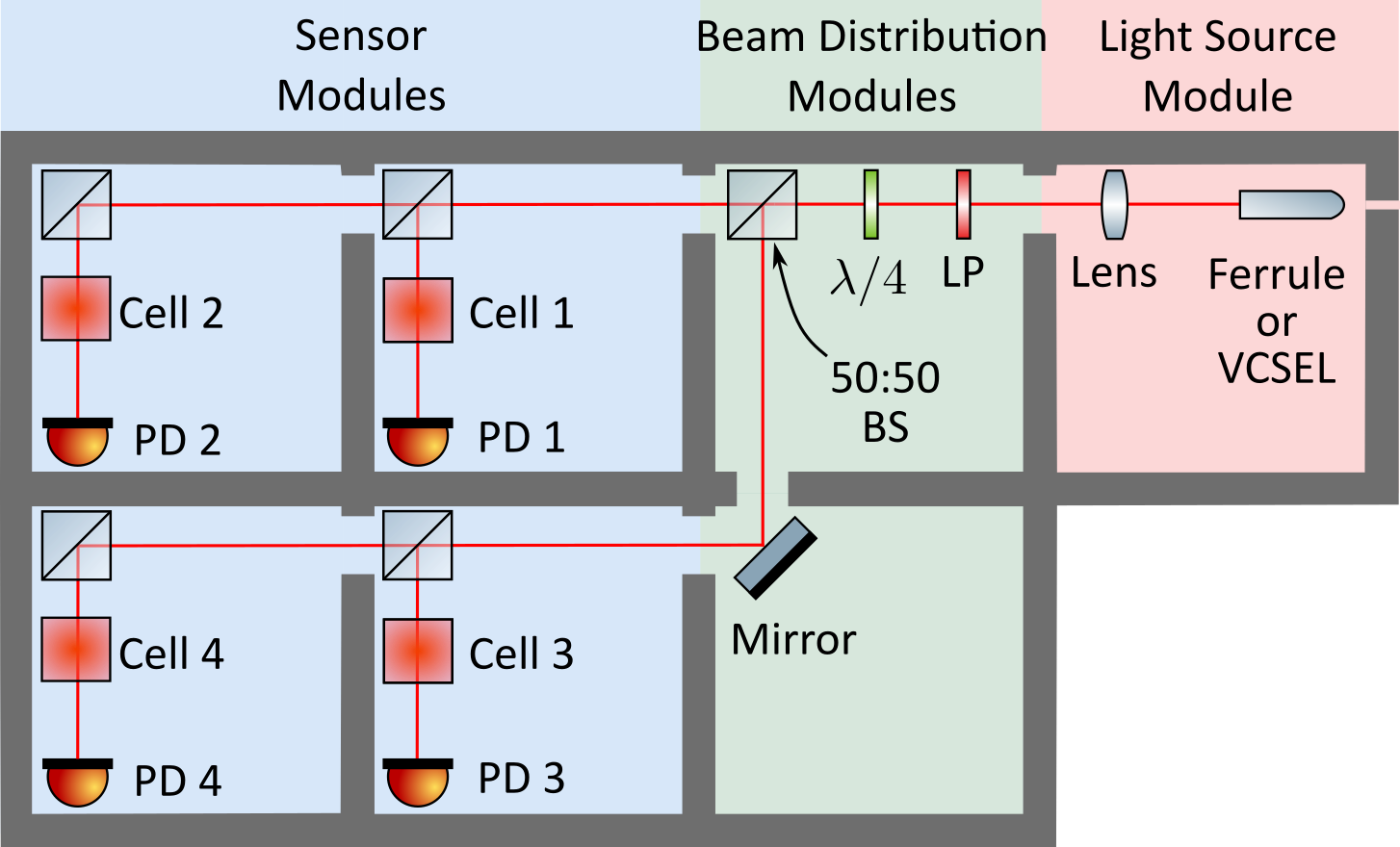}
        \caption{An example of a multi-axis magnetic gradiometer connecting 4 sensor modules in a $2\times 2$ array configuration with light source and beam distribution modules. The laser beam is conditioned with a lens, a linear polariser (LP) and a quarter-waveplate ($\lambda/4$). After passing through vapour cells, the beam power is monitored with photodiodes (PD). A non-polarising 50:50 beam splitter (50:50 BS) and a mirror ensure that the probe light from a single beam is delivered to all sensors.}
        \label{fig:2x2}
   \end{figure}
    
    To illustrate the adaptability of the magnetometer system, two configurations of the instrument, each operating in the ultra-sensitive SERF regime, are presented. A single-sensor configuration was used to capture alpha-rhythms from the brain of a human participant, demonstrating suitability for high sensitivity measurements. Additionally, a two-sensor gradiometer, with superior sensitivity performance through common-mode noise rejection, was characterised.
    Furthermore, we demonstrate that the bandwidth of our system gives it an improved ability to measure higher frequency signals generated  by a spinal cord phantom compared with a commercially available OPM.

    \section{Method} \label{sec.Method}
    Our system utilises \qtyproduct{45 x 45 x 40}{\milli\metre} cuboidal modules, which interconnect to produce a desired OPM system layout. An example of a $2\times 2$ sensor array is shown in Fig.~\ref{fig:2x2}.
    The enclosure of each module is 3D-printed from polycarbonate, allowing for design flexibility whilst also providing excellent mechanical rigidity with a \SI{139}{\degree C} glass transition temperature~\cite{Shamim2014-zu}.
    The system is assembled from three types of modules: light source, sensor, and beam distribution modules. 
    
    The light source modules introduce and condition laser light for use in the magnetometer system. Light can be internally generated by a vertical-cavity surface-emitting laser (VCSEL), or routed from external sources via optical fibres. 
            \begin{figure}
            \centering
            \includegraphics[width=1\columnwidth]{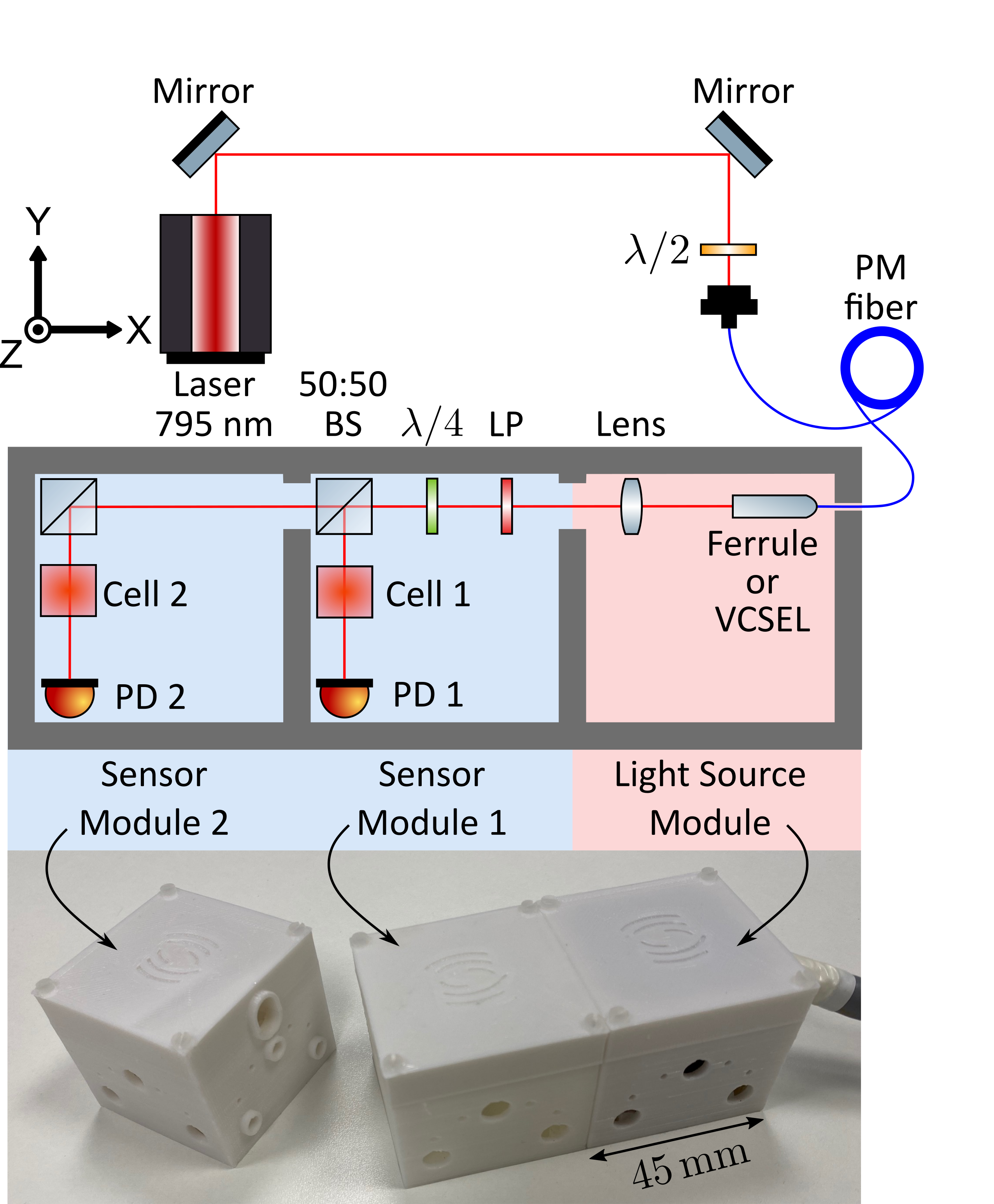}
            \caption{
            A schematic (top) and a photograph (bottom) of the single-axis gradiometer modular OPM consisting of a light source module and two sensor modules. Modules can be snapped together to produce different array configurations.
            }
            \label{fig:optics}
        \end{figure}  
    The sensor modules contain optical elements, vapour cells, magnetic field coils and photodiodes required for magnetic field measurements.
    Finally, beam distribution modules are used for light delivery, allowing for greater flexibility in positioning the sensors.
    Due to the adaptability of its optics packages, measurement schemes and array configurations, the magnetometer can be tailored to a large variety of applications. 
    
    The sensor module design presented in this paper operates in the high-sensitivity SERF regime, which requires large atomic densities in a low magnetic field environment~\citep{allred_high-sensitivity_2002}.
    A single-axis gradiometer (Fig.~\ref{fig:optics}) is constructed using one light source module and two sensor modules in a similar configuration to that in Ref.~\cite{nardelli_conformal_2020}, and can be extended to a multi-axis gradiometer instrument as shown in Fig.~\ref{fig:2x2}.
    The SERF sensor modules utilise the Hanle resonance (zero-field resonance) as a measurement scheme~\cite{Dupont-Roc1969-yj}, relying on the monitoring of the transmitted power of a resonant circularly polarised laser beam probing a vapour cell.
    
    When a magnetic field component that is perpendicular to the laser beam direction, $B_{x}$ (respectively $B_{z}$), is scanned through zero field, atoms are pumped into a dark state and the magnitude of the transmitted light follows a Lorentzian distribution with peak transmission at $B_{x}=0$ (respectively $B_{z}=0$). By applying a modulating field $B_{x,RF}\cos{(\omega t)}$ across the zero-field resonance peak, and passing the photodiode response through a lock-in amplifier (LIA), a dispersion signal is produced, of the form~\cite{cohen-tannoudji_diverses_1970,tierney_optically_2019}
    \begin{equation}
        V(B_x)= A\frac{\gamma B_x \tau}{1 +(\gamma B_x \tau)^2},
    \end{equation}
    where $\gamma$ is the gyromagnetic ratio for the active atomic species, $\tau$ is the relaxation time of the spin-polarised atomic population, and $A$ is a proportionality constant (with units \SI{}{\volt/\pico\tesla}) determined by calibrating the sensor output.
    The LIA also reduces 1/f noise, improving the signal-to-noise ratio of the resonance signal~\cite{kastler_hanle_1973}. A similar argument can be made for the other orthogonal component $B_z$.
    
    Each sensor module contains three orthogonal coil pairs positioned around a vapour cell, which are used to apply offset and modulation fields. The field $B_{x}$ is modulated at frequency $\omega=2\pi\times\SI{926}{\hertz}$ with a peak-to-peak of \SI{60}{\nano\tesla} at the centre of the vapour cell.
    To achieve sufficient atomic density for the SERF regime, the vapour cells are heated using alternating currents at \SI{81}{\kilo\hertz} and \SI{200}{\kilo\hertz}, driven through twisted manganin heating wires wrapped around the vapour cells. These AC frequencies are chosen to be orders of magnitude higher than the rubidium atomic response bandwidth to avoid broadening of the measured resonance. The vapour cell housing is thermally insulated to minimise the required heating power and the surface temperature of the sensor modules. In so doing, a stand-off distance of \SI{8.5}{\milli\metre} from the end of the sensor module housing to the cell centre was achieved. 
    
    Light from an external-cavity diode laser, tuned to the \Rb{} D1 transition (\SI{795}{\nano\metre}), is coupled into a polarisation-maintaining (PM) optical fibre and input to the light source module of the sensor (Fig.~\ref{fig:optics}).
    The linearly polarised beam is then collimated to a $1/e^2$ radius of \SI{1.5}{\milli\metre} before being circularly polarised and split with a 50:50 non-polarising beamsplitter cube.
    The reflected beam is directed through a \qtyproduct{5 x 5 x 5}{\milli\metre} cubic glass \Rb{} cell in sensor module 1, containing \SI{50}{\torr} of N$_{2}$ and \SI{650}{\torr} of Ne as buffer gases \footnote{\qtyproduct{5 x 5 x 5}{\milli\metre} cubic glass \Rb{} cells containing \SI{50}{\torr} of N$_{2}$ and \SI{650}{\torr} of Ne were procured from Twinleaf LLC, USA.}. In sensor module 2, an identical vapour cell is probed with the beam coming from the transmitted arm of the beamsplitter. Photodiodes are placed behind the vapour cells to monitor the probing beam power. Their signals are passed through external transimpedance amplifiers and fed to a pair of synchronised LIAs before being recorded with a digital acquisition card (DAQ) \footnote{LabJack T7 pro, LabJack Corporation, USA.}.
    
    Sensitivity measurements were taken inside a four-layer mu-metal cylindrical shield \footnote{MS-2 mu-metal shield, Twinleaf LLC, USA.} with an internal magnetic field magnitude below \SI{10}{\nano\tesla} and magnetic noise of \SI{10}{\femto\tesla/\sqrt{\hertz}}. The offset fields were applied using the sensor module's internal coils, and modulation fields were generated using field coils built into the shield. The LIA output voltages were calibrated by applying well-defined $B_x$ fields. 
    
    To demonstrate the suitability of the modular OPM system for MEG applications, a single-sensor magnetometer configuration was operated to capture human brain alpha rhythms~\cite{Cohen1968-nf} which occur in the \SIrange{8}{12}{\hertz} band. 
    The experiment was conducted in accordance with the Declaration of Helsinki Ethical Principles, and was approved by the Brighton and Sussex Medical School Research Governance and Ethics Committee (ER/BSMS3100/1). The participant gave written informed consent to take part after explanation of the procedure and purpose of the experiment.
    
     The MEG measurements were taken inside a three-layer mu-metal cylinder, with an internal diameter of \SI{50}{\centi\metre} and axial length of \SI{100}{\centi\metre}.     The sensor setup is similar to the three-module gradiometer configuration shown in Fig. \ref{fig:optics}, with the difference that only one sensor module is present.
     The single-sensor-module OPM was placed at the centre of the cylinder, supported by a wooden arm, against which the participant was able to rest their head. A prerecorded voice prompted the participant to open and close their eyes every \SI{12}{\second} to stimulate an alpha-band response when their eyes were shut (photic blocking)~\cite{Cohen1972-jw}. The optimum placement of the sensor over the primary visual cortex (Oz position) was measured according to the standard 10-10 system~\cite{Koessler2009-pc}. The analogue signal from the OPM-MEG system was sampled at \SI{200}{\hertz} using the same DAQ setup as previously described.

    Finally, the system's temporal resolution enables the capture of higher frequency biomagnetic phenomena. To demonstrate this, a phantom spinal cord signal was generated to form a comparison between the modular OPM system and a commercial OPM.
    The phantom signal was created using WebPlotDigitizer~\cite{Rohatgi2022} based on signals recorded in Ref.~\cite{miyano_visualization_2020} using a SQUID array with a bandwidth of \SI{12}{\kilo\hertz} ~\cite{adachi_recent_2017}.
    For the extracted signal, frequencies up to \SI{650}{\hertz} contain approximately \SI{90}{\percent} of the phantom signal's power.
    The comparison was undertaken inside the four-layer mu-metal cylindrical shield with the phantom signal created using the internal coils of the shield.
    Magnetic field measurements were acquired sequentially from a single-channel modular OPM system and a commercial OPM~\footnote{QZFM Gen. 1, QuSpin Inc., USA.}.
    The phantom spinal cord signal was scaled to ensure the signal was well above the noise floor for both OPM systems, resulting in a peak-to-peak of approximately \SI{4}{\nano\tesla}.
    The analogue outputs of the commercial sensor and modular OPM LIA were recorded using the aforementioned DAQ with a sample frequency of \SI{40}{\kilo\hertz}. The recorded data were filtered with the same notch filters to encompass the first three harmonics of the modulation frequency for the modular system (\SI{926}{\hertz}) and the commercial system (\SI{923}{\hertz}).
    
\section{Results and discussion}
    \begin{figure}
            \centering
            \includegraphics[width=\columnwidth]{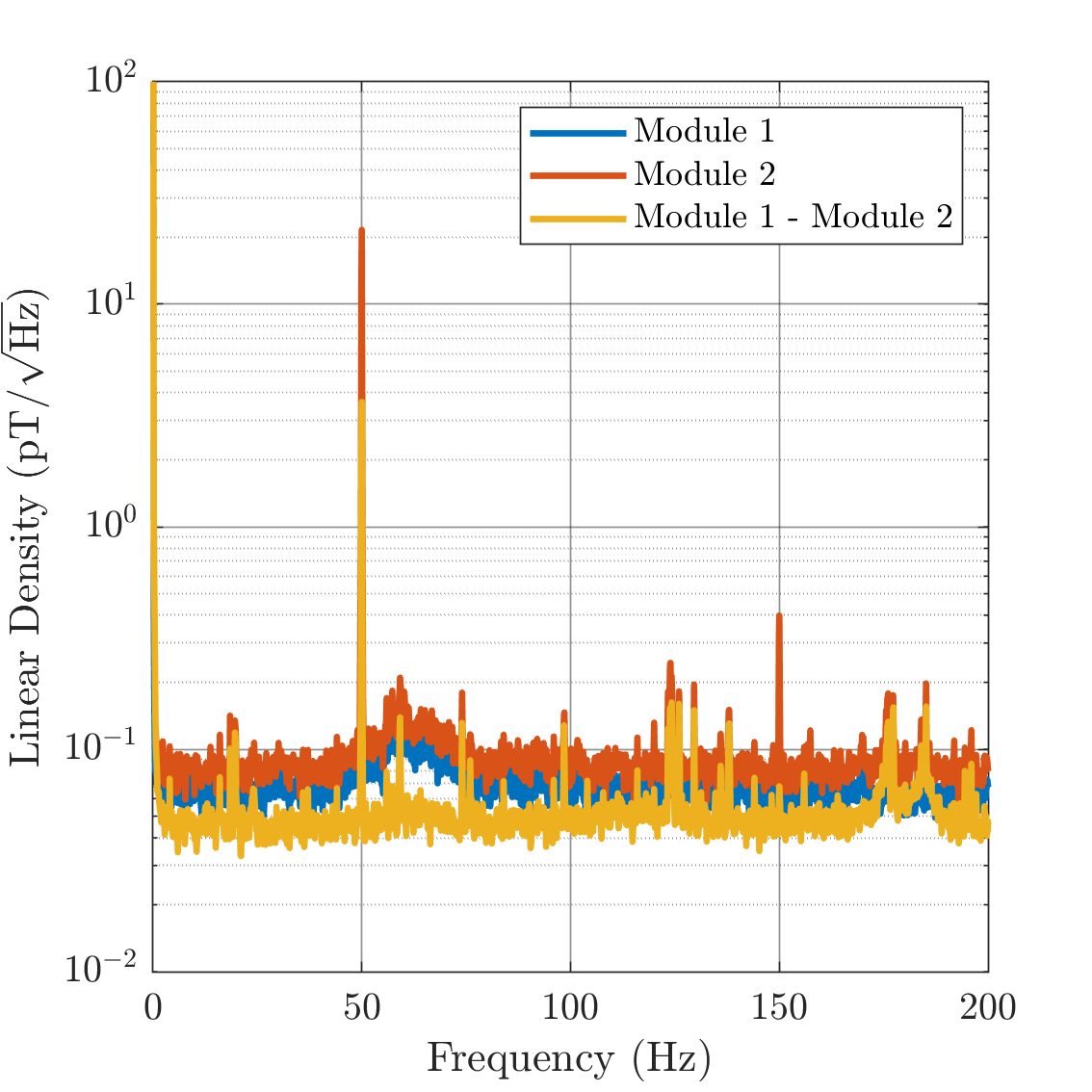}
            \caption{The \SIrange{1}{200}{\hertz} frequency-compensated linear spectral density of the demodulated signals from sensor module 1 (blue), sensor module 2 (red) and from the balanced photodiodes gradiometer configuration (yellow). The gradiometer configuration removes common-mode noise between the two sensors, which can be seen by the noise reduction at \SI{50}{\hertz}, and in the broader peak in the \SIrange{55}{75}{\hertz} band. Sensor module 1 has a noise-floor of \SI{65}{\femto\tesla/\sqrt{\hertz}}, sensor module 2, \SI{83}{\femto\tesla/\sqrt{\hertz}}, and the gradiometer configuration \SI{47}{\femto\tesla/\sqrt{\hertz}} in the \SIrange{5}{45}{\hertz} band.
            }
            \label{fig:ffts}
        \end{figure}
        
  The sensitivity measurements were taken over a period of \SI{60}{\second} during which the magnetometer signals from sensor modules 1 and 2 were recorded.
  Data were analysed for each sensor separately, and as a software gradiometer (sensor module 1 minus sensor module 2).
   The frequency spectra in Fig.~\ref{fig:ffts} show the noise floors of sensor modules 1 and 2 to be \SI{65}{\femto\tesla/\sqrt{\hertz}} and \SI{83}{\femto\tesla/\sqrt{\hertz}}, respectively, in the \SIrange{5}{45}{\hertz} band.
   
\begin{figure}
    \centering
    \includegraphics[width=\columnwidth]{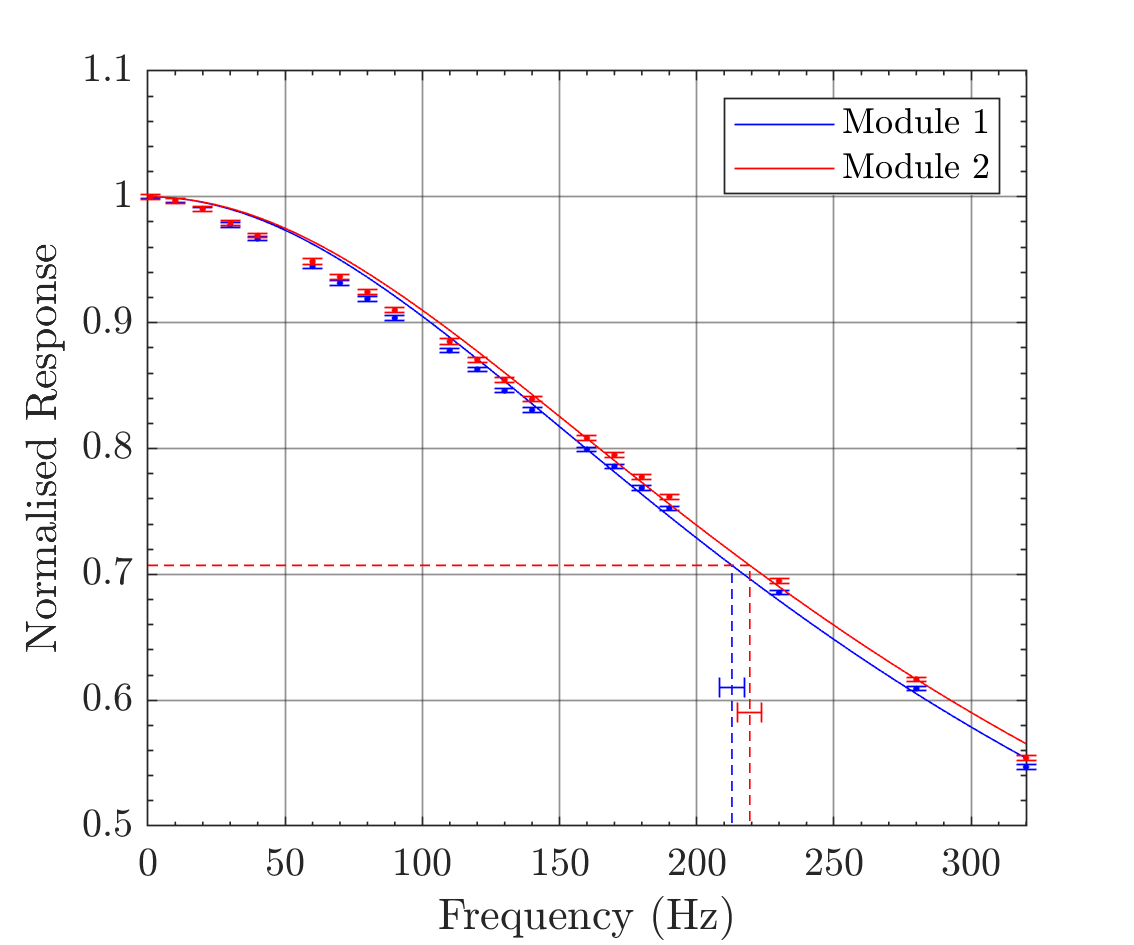}
    \caption{The magnetometer response amplitudes for sensor modules 1 (blue) and 2 (red) as a function of applied signal frequency. The data points are fitted to the response of a single order low pass filter. Dashed lines indicate a \SI{-3}{\decibel} bandwidth of \SI{213(5)}{\hertz} for module 1, and \SI{219(4)}{\hertz} for module 2.}
    \label{fig:f_response}
\end{figure}

   As the response of the \Rb{} atoms and the LIA low-pass filter both limit bandwidth, the linear density spectrum has been divided by the transfer function of each sensor module~\cite{colombo_four-channel_2016}. 
   By applying an oscillating signal of constant amplitude using the internal coils (approx \SI{200}{\pico\tesla}) with varying frequency, the transfer function of each sensor module was measured up to \SI{320}{\hertz}. Each transfer function follows a first-order low-pass filter and the bandwidths of the sensor modules were found to be \SI{213(5)}{\hertz} for module 1, and \SI{219(4)}{\hertz} for module 2, as shown in Fig.~\ref{fig:f_response}.
   
   For this measurement, the \Rb{} vapour cells were operated at temperatures of \SI{129}{\degree C} and \SI{124}{\degree C} in sensor modules 1 and 2, respectively.
   We note that the optimal sensitivity of \Rb{} vapour in the SERF regime occurs at higher temperatures ~\cite{Shah2007-sa}, for example, an increase of the cell temperature to \SI{150}{\degree C} corresponds to a factor-4 increase in vapour density~\cite{Seltzer2008-kh}. Implementing this would, therefore, require an alternative to the printed polycarbonate material. 
   In addition to the potential sensitivity increase from the \Rb{} density, further gains could be made in the low-frequency domain by implementing active feedback on vapour cell temperature and laser power.

\begin{figure}
        \centering
        \includegraphics[width=\columnwidth]{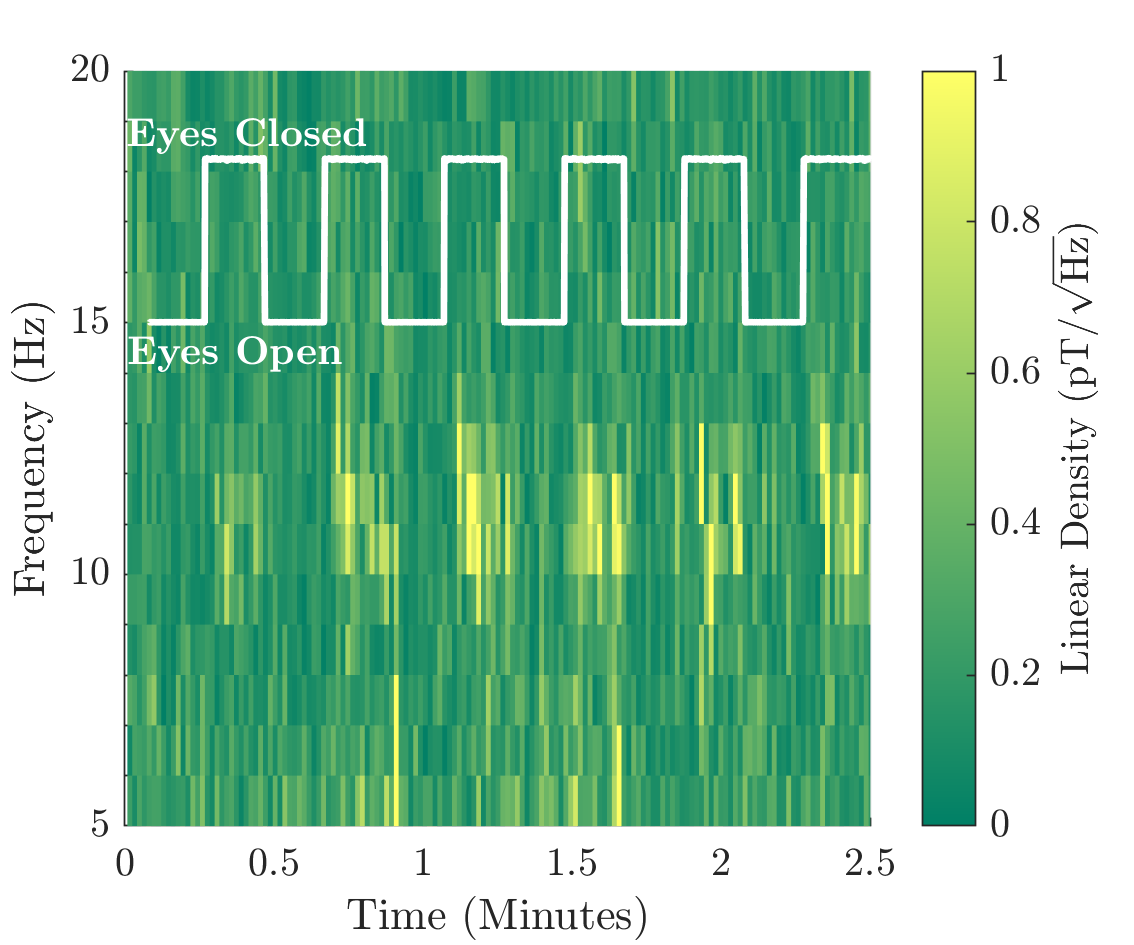}
        \caption{Spectrogram of the \SIrange{8}{12}{\hertz} alpha-band response to a participant opening/closing their eyes. The overlaid white trace denotes the audio cue trigger signal from the stimulus PC. High values of the trigger signal indicate the time when the participant is instructed to close their eyes, the low value is when the participant has their eyes open. The spectrogram colour scheme details the high periods of activity in yellow, and the low periods in green. A peak in activity of \SI{1}{\pico\tesla/\sqrt{\hertz}} is observed in the the \SIrange{8}{12}{\hertz} region during the third eyes-closed time period.}
        \label{fig:spectrogram}
    \end{figure}

   Data from the alpha rhythm measurement with a single sensor module were analysed in the frequency domain, with the spectrogram shown in Fig.~\ref{fig:spectrogram}. We see an increase in the activity in the \SIrange{8}{12}{\hertz} alpha-band when the participant's eyes are closed, with a strong correlation with the TTL cue. Although we already show that the current modular sensor is suitable for MEG, future improvements of the signal-to-noise ratio can be made by suppressing common-mode noise using a gradiometer. 
  
    Larger bandwidths were observed for the modular OPM system compared with the commercial OPM which is approximately \SI{135}{\hertz}~\cite{osborne_fully_2018}. Higher OPM bandwidths can be achieved at the expense of sensitivity by broadening the width of the resonance~\cite{tierney_optically_2019}, which is dependent on multiple factors, including the rate of optical pumping, spin destruction collisions, collisions with the vapour cell wall and the presence of magnetic field gradients~\cite{shah_spin-exchange_2009}. Alternatively, the sensitive bandwidth range can be shifted to higher frequencies by applying DC field offsets~\cite{savukov_high-sensitivity_2017}.
    This capability is important for resolving higher frequency biomagnetic phenomena. 
    
\begin{figure}[h!]
        \centering
        \includegraphics[width=\columnwidth]{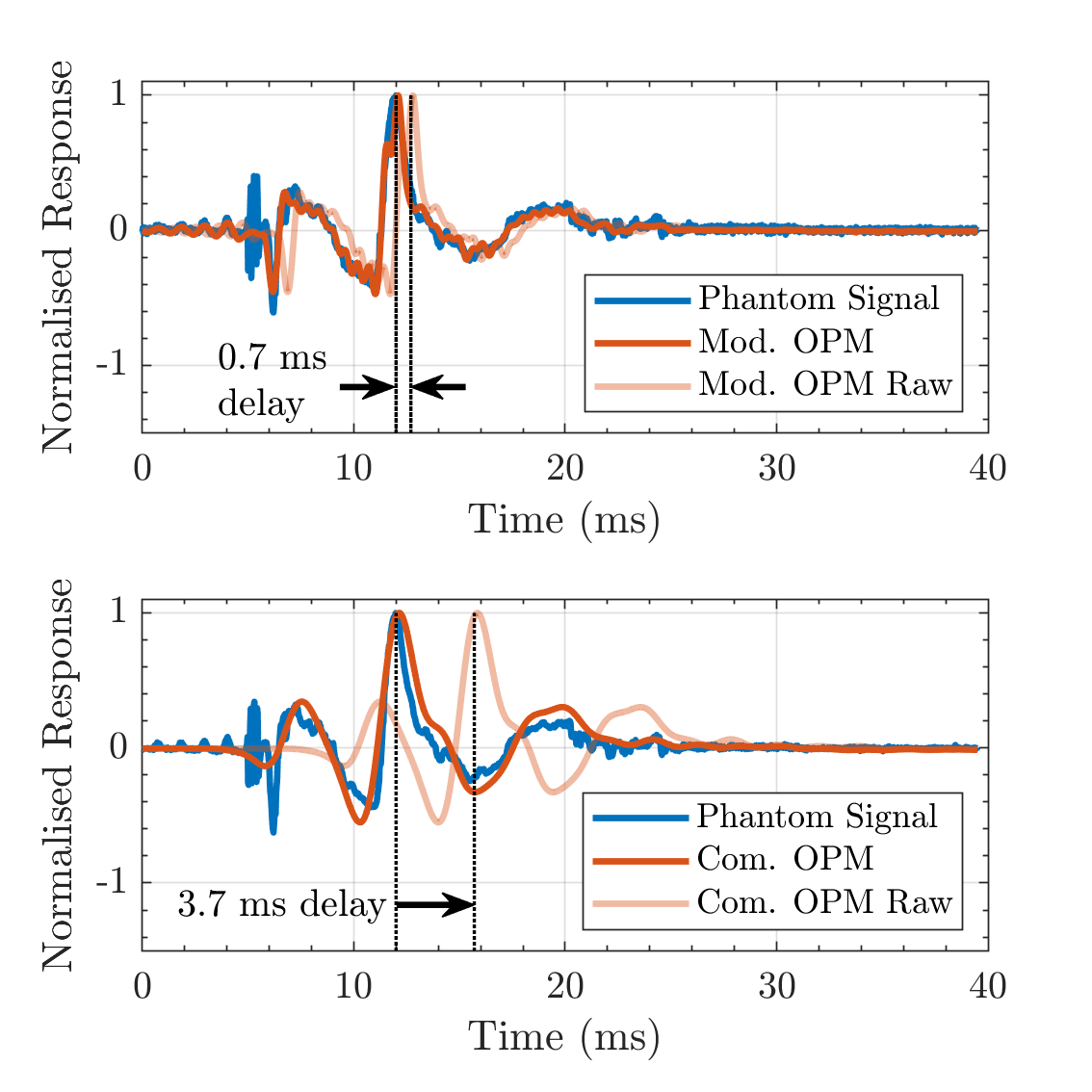}
        \caption{A comparison in the responses of the modular (top) and commercial (bottom) OPM systems to the phantom spinal cord signal. 
        Time delays were optimised using RMSD minimisation with respect to the raw phantom trace. The uncompensated raw data are shown as faded traces.
        }
        \label{fig:comparison}
    \end{figure}

    In this case we present the advantages of increased bandwidth on the measurement of spinal cord activity with a phantom signal.
    The extracted spinal cord signal (as described in the Method section) is used to drive the shield's internal coil. The responses to this signal as recorded by the modular and commercial OPM systems are presented in Fig.~\ref{fig:comparison}.
    Time delays were optimised using root mean square deviation (RMSD) minimisation with respect to the raw phantom trace and signal amplitudes normalised to the peak at \SI{20}{\milli\second}. Time delays of \SI{0.7}{\milli\second} for the modular OPM and \SI{3.7}{\milli\second} for the commercial OPM were observed.
    We believe this delay is characteristic of each system's signal processing chain.
    The modular OPM system provides a more accurate replication of the signal trace, with a RMSD of 0.05 compared to the commercial sensor's RMSD of 0.10. The features of the high-frequency phantom signal around \SI{12}{\milli\second} are not well resolved by the commercial OPM.

    We have designed a modular OPM system and demonstrated its operation in single and dual channel configurations. We have also utilised the system to record MEG signals from a human participant and demonstrated an improved bandwidth over a commercial OPM system in resolving high-frequency features from a phantom spinal cord signal.
    The ability of this highly customisable OPM system to implement new measurement schemes, array configurations, dual-scheme OPM arrays, or multi-modal quantum sensor systems, offers a unique combination of adaptability and capability for in-field use.
    This modular system is an important development tool for the advancement of dense OPM arrays in medical and industrial fields, providing a rapid testing platform to benchmark OPM system architectures.
  
\section*{Data Repository}
    The raw data to support the findings of this study are available in Ref.~\cite{coussens_modular_data}.

\section*{Acknowledgements}
    This work was supported by the UK Quantum Technologies Hub for Sensors and Timing (EPSRC Grant EP/T001046/1), University of Sussex Strategic Development Fund and Innovate UK: Batteries - ISCF 42186 Quantum sensors for end-of-line battery testing.
    
\section*{CRediT Author Statement}
    
    \textbf{T.~Coussens}: Conceptualisation, Methodology, Investigation, Visualisation, Writing – original draft, Writing – review \& editing.
    \textbf{A.~Gialopsou} : Investigation, Visualisation, Writing – review \& editing.
    \textbf{C.~Abel}: Methodology, Writing – review \& editing.
    \textbf{M.~Bason}: Methodology, Writing – review \& editing.
    \textbf{T.~James}: Visualisation, Writing – review \& editing.
    \textbf{W.~Evans}: Writing – review \& editing.
    \textbf{M.~Woodley}: Writing – review \& editing.
    \textbf{D.~Nightingale}: Writing – review \& editing.
    \textbf{D.~Nicolau}: Writing – review \& editing.
    \textbf{L.~Page}: Writing – review \& editing.
    \textbf{P.~Krüger}: Writing – review \& editing, Supervision, Funding acquisition.
    \textbf{F.~Oru\v{c}evi\'{c}}: Writing – review \& editing, Supervision, Funding acquisition.
    
\newpage

\bibliographystyle{ieeetr}  
\bibliography{refs2, refs}

\end{document}